\documentclass[11pt]{article}

\usepackage[utf8]{inputenc}
\usepackage[T1]{fontenc}
\usepackage{amsmath}
\usepackage{amssymb}
\usepackage{amsfonts}
\usepackage{bm}
\usepackage{graphicx}
\usepackage{booktabs}
\usepackage{array}
\usepackage{longtable}
\usepackage[margin=1in]{geometry}
\usepackage{pdflscape}
\usepackage[colorlinks=true,linkcolor=blue,citecolor=blue,urlcolor=blue]{hyperref}
\usepackage{url}
\usepackage{enumitem}
\usepackage{authblk}

\title{Decentralised Consensus Learning Networks: SME Rotation Without Centralised Reward}
\author{Florin Neagu}
\date{}

\begin{document}

\maketitle

\begin{abstract}
Centralised reward signals dominate the design of learning systems in AI, yet they impose a single external arbiter on what constitutes correct or valuable knowledge. This paper presents an alternative: a decentralised, consensus-based multi-agent learning network in which expertise emerges from peer validation rather than prescribed reward. Agents update beliefs through weighted social consensus, with trust allocated in proportion to competence --- itself inferred from peer consistency rather than ground truth. Subject-matter expert (SME) status is assigned dynamically as a top-percentile competence rank, not a fixed label. We report results from eighty-four agent-based simulation runs spanning N=30 to N=10,000 agents, two graph topologies, a sparse large-scale implementation, a systematic investigation of an oligarchy-adjacent failure mode and its resolution, a Phase 2 series extending beliefs from scalars to D=5 unit vectors on the hypersphere, and a Phase 3 series characterising multi-seed robustness, long-run convergence behaviour, a dimensionality sweep across D={1,2,5,10,20}, and a high-dimensional equilibrium study across D={20,50,100,125,150,175,200,500}. The Phase 1 central finding is that SME rotation is robust, persistent, topology-invariant, and scale-invariant: 90--100\% of agents held SME status at some point across all primary runs and across a 9-seed robustness check (Series A0-scalar), 67--86\% (mean $\approx$77\%) of all expertise turnover occurred after beliefs had fully converged across that same multi-seed check, and this post-convergence rotation strengthened rather than weakened as network size increased from N=120 to N=10,000. The Phase 2 and Phase 3 Series A0--A2 central finding is that vector beliefs produce convergence timescale heterogeneity with cascade dynamics, and that belief dimensionality D induces five qualitatively distinct dynamical regimes with genuine phase structure. The Phase 3 Series A3 finding is that the (D, ETA) parameter space contains a high-D operating zone, tested across D=150--200 at six seeds (the original seed=42 plus five additional seeds, Series A3c), where the network stabilises with strong but incomplete consensus: five of six seeds cluster tightly in the eq\_cosim $\approx$ 0.76--0.83 range, with higher D correlating with lower consensus overall (one seed shows a mild non-monotonic step, discussed in Section 4.9) --- the sixth seed is a wider, structurally-explained outlier (an unusually low-degree node, independently verified) rather than a break in the trend. This equilibrium is accompanied by a monotone, unavoidable trade-off: as D increases and belief diversity increases, SME expertise concentration also increases, with a single agent typically holding top status the large majority of timesteps at high D. This is reinterpreted not as a failure but as emergent behaviour consistent with the model's design intent: in a high-dimensional near-consensus state, the agent most consistently aligned with the complex multi-dimensional consensus direction is correctly identified as the expert, and the seed-to-seed variability in exactly how concentrated that expertise becomes reflects genuine path-dependence in how the network settles, not measurement noise. The ETA sensitivity study confirms that this expertise concentration is dimensionality-driven, not noise-driven --- reducing ETA shifts the equilibrium cosim toward 1.0 without recovering rotation health, falsifying the hypothesis that ETA and D are independent design levers for diversity and rotation respectively.
\end{abstract}

\section{Motivation}

The prevailing paradigm for training learning systems relies on a centralised reward signal: an external function that evaluates outputs and drives parameter updates toward higher-scoring behaviour. This approach has produced remarkable results, but it carries structural assumptions that may limit its applicability to certain classes of problems. The reward function must be specified in advance, must remain stable over the learning period, and must be capable of evaluating all relevant outputs. In domains where knowledge is contested, multi-dimensional, or rapidly evolving --- much of what human experts actually do --- these assumptions are difficult to satisfy.

Human knowledge communities do not, in general, operate on centralised reward. Academic fields, engineering teams, and professional communities develop and validate expertise through a distributed process: ideas are proposed, debated, adopted, or discarded based on peer response. An expert is not someone who received a high score from an external oracle; an expert is someone whose views are consistently sought out and weighted heavily by peers. Expertise in this sense is conferred by the community, not assigned by a designer.

This observation motivates the model described here. Rather than asking how a learning system can maximise a prescribed objective, we ask: under what conditions does peer-validated expertise emerge, rotate, and remain broadly distributed in a network of agents who learn only from each other? The question is conceptual and the model is a proof of concept, not a production system. We make no claim that this architecture is ready for deployment or that it outperforms supervised learning on any practical benchmark. The contribution is narrower: to demonstrate that the structural conditions for rotating, non-oligarchic expertise emergence can be satisfied in a simple agent-based model without any centralised reward signal.

There is a deeper implication that the Phase 2 and Phase 3 results make visible. A learning system trained to minimise disagreement with human raters --- the dominant paradigm in contemporary AI --- is structurally operating in a high social learning rate (ALPHA), low innovation rate (ETA) regime. Such a system converges reliably toward the human corpus. In the language of this model, its belief vectors collapse toward the attractor defined by the training signal, and once consensus is reached, genuine novelty generation requires perturbations large enough to overcome the social pull. The model predicts that this is structurally difficult: the social term is correlated across the neighbourhood while the innovation term is independent per agent, so at any reasonable scale the social pull dominates. The result is a ceiling on innovation that is not a tuning failure but a property of the architecture's fixed-point structure.

The Phase 3 long-run results add a further observation: convergence in high-dimensional belief space is not merely slow --- it is highly heterogeneous and cascade-driven. In a network operating with vector beliefs, individual topics find their own convergence timescales independently, driven by initial geometry. Some ideas tip rapidly; others wander for thousands of steps before cascading; others may not tip within any practical operational window and are effectively dropped by that cluster. This is the positive complement to the centralised reward ceiling observation: a network with vector beliefs and balanced ALPHA/ETA dynamics naturally produces the selective idea propagation that characterises healthy knowledge ecosystems --- where not every idea reaches every community, and the ideas that do propagate earn their consensus through local alignment rather than global mandate.

The Series A2 dimensionality sweep establishes that D --- the number of dimensions in the belief vector --- is a regime-determining design lever, not a neutral parameter. The Series A3 high-dimensional equilibrium study extends this finding: at sufficiently high D, the network enters a Pareto-efficient operating zone where strong broad agreement coexists with meaningful residual diversity, reached in tens of steps and maintained indefinitely. This zone has no analogue in scalar belief networks or in centralised reward architectures, which either collapse to full consensus or require explicit diversity-preservation mechanisms to avoid it. Here diversity and consensus coexist as a structural property of the geometry.

The model draws on four bodies of prior work. Holland's (1995) account of complex adaptive systems establishes the broader theoretical basis for this paper's central premise: that robust, non-random order --- including the rotating, non-oligarchic expertise structures examined here --- can emerge from local interaction rules without any centralised designer or reward signal. DeGroot's (1974) consensus model establishes the mathematical foundations of belief averaging in social networks. Deffuant et al.'s (2000) bounded confidence framework introduces the insight that agents update only from peers whose beliefs are sufficiently similar to their own, preventing runaway consensus under certain conditions. Watts and Strogatz's (1998) small-world network model provides the graph topology used in the primary simulation runs, chosen because its combination of local clustering and short average path lengths approximates the structure of real social and professional networks.

\section{Proposed Model}

The model defines a network of $N$ autonomous agents (nodes) connected by an undirected graph $G = (V, E)$ with adjacency matrix $A = [a_{ij}]$. Each agent maintains beliefs and competence scores across $M$ topics. For agent $i$ on topic $k$ at time $t$:

\begin{itemize}
\item $c_i(k,t) \in [0,1]$ --- competence score
\item $w_{ij}(k,t)$ --- trust weight toward neighbour $j$ on topic $k$, normalised so $\sum_j w_{ij} = 1$
\end{itemize}

Belief representation differs between phases. In Phase 1, $b_i(k,t) \in [0,1]$ is a scalar. In Phase 2 and Phase 3, $\mathbf{b}_i(k,t) \in \mathbb{R}^D$ is a unit vector on the D-dimensional hypersphere, with $\|\mathbf{b}_i\| = 1$ enforced after each update by normalisation. D is the belief dimensionality --- the number of independent aspects of a topic that the belief vector simultaneously encodes. The Series A2 sweep tested D $\in \{1, 2, 5, 10, 20\}$ and found qualitatively distinct dynamical regimes at each value (Section 4.8). Series A3 extended the sweep to D $\in \{20, 50, 100, 125, 150, 175, 200, 500\}$ to characterise the high-D equilibrium regime (Section 4.9).

\subsection{Belief Update}

\textbf{Phase 1 (scalar):}

\[
b_i(k,t+1) = (1 - \alpha - \eta)\, b_i(k,t) \;+\; \alpha \sum_j w_{ij}(k,t)\, b_j(k,t) \;+\; \eta\, u_i(k,t)
\]

where $u_i(k,t) \sim \text{Uniform}(-0.3, 0.3)$, clipped so that $b_i$ remains in $[0,1]$.

\textbf{Phase 2 and Phase 3 (vector):}

\[
\mathbf{b}_i(k,t+1) = \text{normalise}\!\left[(1 - \alpha - \eta)\, \mathbf{b}_i(k,t) \;+\; \alpha \sum_j w_{ij}(k,t)\, \mathbf{b}_j(k,t) \;+\; \eta\, \boldsymbol{\epsilon}_i(k,t)\right]
\]

where $\boldsymbol{\epsilon}_i(k,t) \sim \mathcal{N}(\mathbf{0}, \sigma^2 \mathbf{I}_D)$ with $\sigma = 0.1$. The normalisation step means innovation noise acts as angular rotation. The effective angular amplitude scales as $\sigma\sqrt{D}$ --- at D=5 approximately 0.224 before normalisation; at D=175 approximately 1.322. This scaling drives the regime transitions observed in Series A2 and the Pareto equilibrium behaviour in Series A3.

The parameters $\alpha$ (social learning rate) and $\eta$ (innovation rate) govern the balance between collective alignment and individual novelty generation. Series A3 established that ETA and D are not independent design levers: at high D, reducing ETA shifts the equilibrium cosim toward 1.0 without recovering SME rotation health. D is the primary control for equilibrium belief diversity; ETA is a secondary timescale modulator whose influence on regime identity is subordinate to D at high dimensionality.

Trust weights are proportional to competence:

\[
\tilde{w}_{ij}(k,t) = a_{ij} \cdot c_j(k,t), \qquad w_{ij}(k,t) = \frac{\tilde{w}_{ij}}{\sum_l \tilde{w}_{il}}
\]

\subsection{Bounded Confidence Filter}

\textbf{Phase 1:} $w_{ij}^{\text{bc}} = 0$ if $|b_i(k,t) - b_j(k,t)| > \epsilon$

\textbf{Phase 2 and Phase 3:} $w_{ij}^{\text{bc}} = 0$ if $\mathbf{b}_i(k,t) \cdot \mathbf{b}_j(k,t) < \kappa$

With $\kappa = 0.0$, all neighbours pass the filter. Run 11 used $\kappa = 0.5$; Runs 12--69 used $\kappa = 0.0$. Run 12 demonstrated that $\kappa = 0.5$ freezes social learning under Gaussian initialisation.

\subsection{Competence Update and SME Emergence}

\textbf{Phase 1:} $\Delta_i(k,t) = -\sum_j w_{ij}^{\text{bc}}(k,t)\; \bigl(b_i(k,t) - b_j(k,t)\bigr)^2$

\textbf{Phase 2 and Phase 3:} $\Delta_i(k,t) = \sum_j w_{ij}^{\text{bc}}(k,t)\; \mathbf{b}_i(k,t) \cdot \mathbf{b}_j(k,t)$

\[
c_i(k,t+1) = (1 - \lambda)\, c_i(k,t) \;+\; \lambda \cdot \hat{\Delta}_i(k,t)
\]

where $\hat{\Delta}$ is $\Delta$ mapped to $[0,1]$ by min-max normalisation across connected agents.

Isolated agents (degree zero) receive $\Delta_i = 0$, which maps to 1.0 under min-max normalisation --- spuriously rewarding non-participation. The two-guard fix (NaN sentinel exclusion + zero override after normalisation) is required for correctness on any graph admitting degree-zero nodes.

SME status: agent $i$ is an SME on topic $k$ at time $t$ if $c_i(k,t)$ exceeds the 90th percentile across all agents on that topic.

\subsection{Trust Evolution and Anti-Oligarchy}

\[
w_{ij}(k,t+1) \propto w_{ij}(k,t) \cdot \exp\!\big(\gamma \cdot \text{sim}(b_i, b_j) + \gamma_c \cdot c_j(k,t) - \mu \cdot w_{ij}(k,t)\big)
\]

Tested in Phase 2 (Run 7, $\mu \in \{1.0, 5.0\}$); showed no effect because the apparent hub dominance was the isolated-node boundary condition, not degree-skew. The Series A3 ETA sensitivity study further established that the high-D expertise concentration is dimensionality-driven and not addressable by noise or trust-penalty parameters --- it reflects genuine multi-dimensional alignment tracking (Section 4.9).

\section{Simulation Design}

Eighty-four simulation runs were executed across three phases.

\begin{landscape}
\footnotesize
\begin{longtable}{>{\raggedright\arraybackslash}p{1.0cm}>{\raggedright\arraybackslash}p{0.6cm}>{\raggedright\arraybackslash}p{0.9cm}>{\raggedright\arraybackslash}p{2.0cm}>{\raggedright\arraybackslash}p{1.6cm}>{\raggedright\arraybackslash}p{1.1cm}>{\raggedright\arraybackslash}p{1.6cm}>{\raggedright\arraybackslash}p{1.3cm}>{\raggedright\arraybackslash}p{5.2cm}}
\toprule
Run & Phase & N & Topology & Filter & $\eta$ & D & T & Primary question \\
\midrule
\endhead
\bottomrule
\endfoot
1 & 1 & 30 & WS k=4, p=0.1 & $\epsilon$=0.30 & 0.05 & 1 (scalar) & 200 & Baseline \\
2 & 1 & 30 & WS k=4, p=0.1 & $\epsilon$=0.15 & 0.15 & 1 & 200 & Tighter filter / higher ETA \\
3 & 1 & 30 & WS k=4, p=0.1 & $\epsilon$=0.15 & 0.15 & 1 & 200 & Stronger noise \\
4 & 1 & 120 & WS k=4, p=0.1 & $\epsilon$=0.15 & 0.15 & 1 & 500 & Scale and post-convergence rotation \\
5 & 1 & 120 & SBM 4$\times$30 & $\epsilon$=0.15 & 0.15 & 1 & 500 & Topology invariance \\
6 & 1 & 10,000 & SBM 500$\times$20 & $\epsilon$=0.15 & 0.15 & 1 & 500 & Large scale \\
7 & 2 & 120 & SBM 4$\times$30 & $\epsilon$=0.15 & 0.15 & 1 & 500 & Trust penalty \\
8 & 2 & 120 & SBM 4$\times$30 & $\epsilon$=0.15 & 0.15 & 1 & 500 & Degree-normalised peer-consistency \\
9 & 2 & 120 & SBM 4$\times$30 & $\epsilon$=0.15 & 0.15 & 1 & 500 & Uniform C0 \\
10 & 2 & 120 & SBM 4$\times$30 & $\epsilon$=0.15 & 0.15 & 1 & 500 & Isolated-node guard \\
11 & 2 & 30 & WS k=4, p=0.1 & $\kappa$=0.5 & 0.15 & 5 & 200 & Vector baseline --- uniform init \\
12 & 2 & 30 & WS k=4, p=0.1 & $\kappa$=0.5 & 0.15 & 5 & 200 & Gaussian init \\
13 & 2 & 30 & WS k=4, p=0.1 & $\kappa$=0.0 & 0.15 & 5 & 200 & BC filter off \\
14 & 2 & 30 & WS k=4, p=0.1 & $\kappa$=0.0 & 0.15 & 5 & 1,000 & Is diversity structural? \\
15--23 & 3 & 30 & WS k=4, p=0.1 & $\kappa$=0.0 & 0.15 & 5 & 500 & Series A0: multi-seed vector robustness (9 seeds) \\
24--32 & 3 & 30 & WS k=4, p=0.1 & $\epsilon$=0.15 & 0.15 & 1 (scalar) & 200 & Series A0-scalar: multi-seed scalar robustness (9 seeds; executed 2026-07-27, retroactively filling a numbering gap reserved since the original 2026-06-18 amendment) \\
33--50 & 3 & 30 & WS k=4, p=0.1 & $\kappa$=0.0 & 0.01--0.15 & 5 & 500 & ETA sweep: 6 $\times$ 3 seeds \\
51--53 & 3 & 30 & WS k=4, p=0.1 & $\kappa$=0.0 & 0.15 & 5 & 5,000 & Long run T=5000 \\
54 & 3 & 30 & WS k=4, p=0.1 & $\kappa$=0.0 & 0.15 & 5 & 20,000 & Ultra-long run seed 42 \\
55 & 3 & 30 & WS k=4, p=0.1 & $\kappa$=0.0 & 0.15 & 1 & 500 & D sweep: D=1 (vector) \\
56 & 3 & 30 & WS k=4, p=0.1 & $\kappa$=0.0 & 0.15 & 2 & 20,000 & D sweep: D=2 \\
57 & 3 & 30 & WS k=4, p=0.1 & $\kappa$=0.0 & 0.15 & 5 & 20,000 & D sweep: D=5 validation \\
58 & 3 & 30 & WS k=4, p=0.1 & $\kappa$=0.0 & 0.15 & 10 & 50,000 & D sweep: D=10 \\
59 & 3 & 30 & WS k=4, p=0.1 & $\kappa$=0.0 & 0.15 & 20 & 100,000 & D sweep: D=20 \\
60--64 & 3 & 30 & WS k=4, p=0.1 & $\kappa$=0.0 & 0.15 & 20--500 & 5,000 & A3a: high-D equilibrium sweep \\
65--67 & 3 & 30 & WS k=4, p=0.1 & $\kappa$=0.0 & 0.15 & 125--175 & 5,000 & A3b sub-run 1: fine D sweep \\
68--69 & 3 & 30 & WS k=4, p=0.1 & $\kappa$=0.0 & 0.05--0.10 & 150 & 5,000 & A3b sub-run 2: ETA sensitivity \\
70--84 & 3 & 30 & WS k=4, p=0.1 & $\kappa$=0.0 & 0.15 & 150--200 & 5,000 & A3c: seed-robustness check for the D=150--200 zone (5 additional seeds $\times$ 3 D values) \\
\end{longtable}
\end{landscape}

All runs: $\alpha = 0.3$, $\lambda = 0.1$, seed 42 unless noted. Phase 1 uses scalar beliefs. Phase 2 and 3 use D-dimensional unit vectors with Gaussian initialisation (Runs 12--69) or uniform positive-orthant initialisation (Run 11).

\section{Results}

\subsection{SME Rotation Is Robust and Topology-Invariant}

Across Runs 1--5 and Run 10, between 93\% and 100\% of agents held SME status at least once (the 93\% floor comes from Runs 1 and 2; Run 10, the isolated-node fix, sits at 97.5\%). 120/120 agents in Run 4 (WS, N=120). 117/120 (97.5\%) in Run 10 (SBM, corrected). Topology change from WS to SBM produced indistinguishable results --- the consensus mechanism drives rotation; graph structure does not.

\begin{figure}[htbp]
\centering
\includegraphics[width=0.95\textwidth]{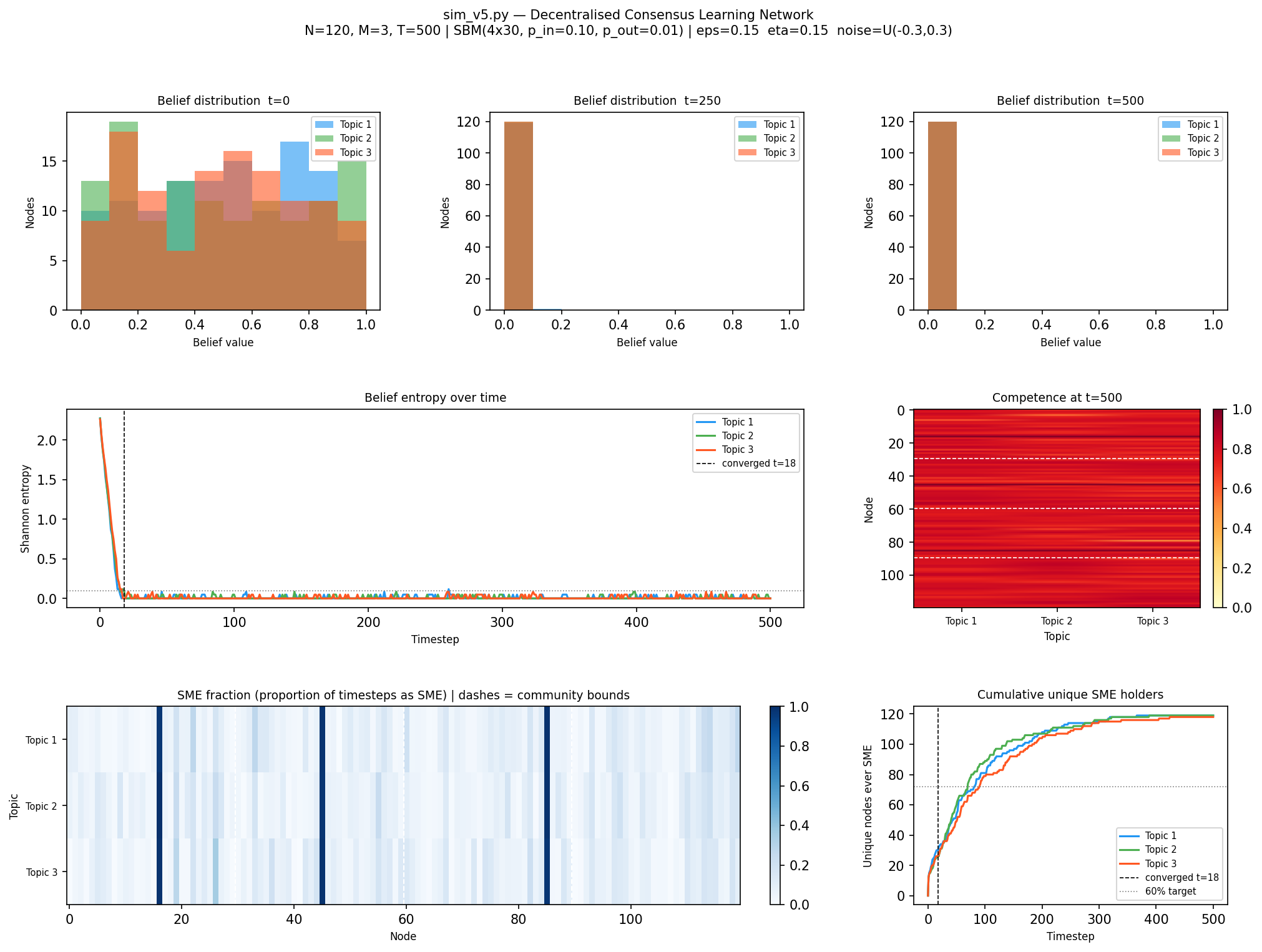}
\caption{Run 5 results (SBM topology, N=120), showing rotation and SME-fraction panels directly comparable to the Watts-Strogatz results of Run 4 (Section 4.1). The consensus mechanism, not graph structure, drives rotation.}
\label{fig1}
\end{figure}

\subsection{Post-Convergence Rotation Separates Belief and Competence Dynamics}

In Run 4 (N=120, T=500), belief entropy collapsed by timestep 17. Despite this:
\begin{itemize}
\item Topic 1: 90/120 agents (75\%) first entered SME after t=17
\item Topic 2: 88/120 agents (73\%) first entered SME after t=17
\item Topic 3: 94/120 agents (78\%) first entered SME after t=17
\end{itemize}

Run 5 replicated: 74--78\% post-convergence. Expertise rotation is driven by competence dynamics, not belief dynamics.

\begin{figure}[htbp]
\centering
\includegraphics[width=0.95\textwidth]{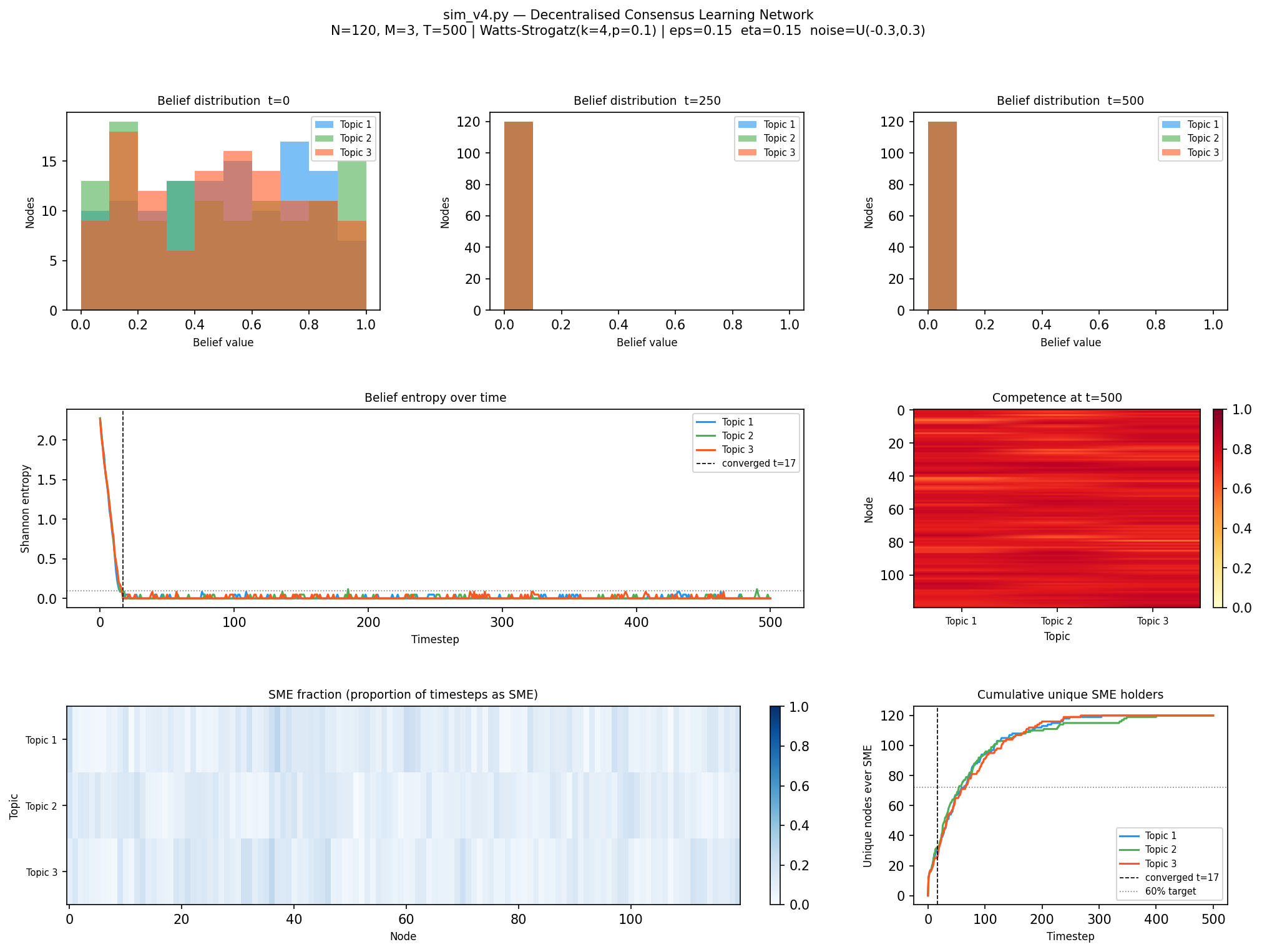}
\caption{Run 4 results (N=120), showing belief entropy collapse at t=17 alongside the per-topic SME-fraction panels underlying the 73--78\% post-convergence figures cited above --- the actual source run for this section's central claim.}
\label{fig2}
\end{figure}

\subsection{Multi-Seed Confirmation of Scalar SME Rotation (Series A0-scalar)}

The scalar-belief results reported above (Sections 4.1--4.2) were, like every other Phase 1 run, based on a single random seed (42). Series A0 (Section 4.7) closed the equivalent gap for the vector-belief (D=5) case; the matching scalar-belief check was planned alongside Series A0 under the same 2026-06-18 amendment, given the run numbers 24--32, but not executed at the time. It was completed retroactively as \textbf{Series A0-scalar}: the same 9 seeds used in Series A0 ({1, 7, 13, 21, 42, 77, 99, 123, 256}), applied to Run 2's exact scalar configuration (N=30, WS k=4 p=0.1, $\epsilon$=0.15, ALPHA=0.3, ETA=0.15, T=200).

Belief convergence occurred in all 9/9 seeds (t\_conv 12--14), consistent with Section 4.4's claim that scalar convergence is a structural property of the update equation rather than a seed- or noise-dependent outcome. Ever-SME participation ranged 90--100\% across seeds and topics (27--30 of 30 agents per topic), consistent with --- though slightly broader than --- the single-seed 93--100\% figure reported in Section 4.1. Post-convergence rotation, measured identically to Section 4.2's per-topic statistic, ranged 66.7--85.7\% across all 27 (seed, topic) observations, with a mean of approximately 76.7\%, closely reproducing the single-seed 73--81\% range reported above. Top-holder concentration ranged approximately 20--39\%, consistent with the 28--38\% range established in Run 10 (Section 4.3).

These results confirm that Phase 1's central rotation claim is not a seed-42 artefact: the qualitative pattern (robust, persistent, post-convergence-dominant rotation) and the approximate quantitative range both reproduce closely across 9 independent seeds. The original single-seed figures sit at the upper end of, rather than outside, the range this multi-seed check produces. The Abstract and Conclusion report the fuller multi-seed range accordingly.

\begin{figure}[htbp]
\centering
\includegraphics[width=0.95\textwidth]{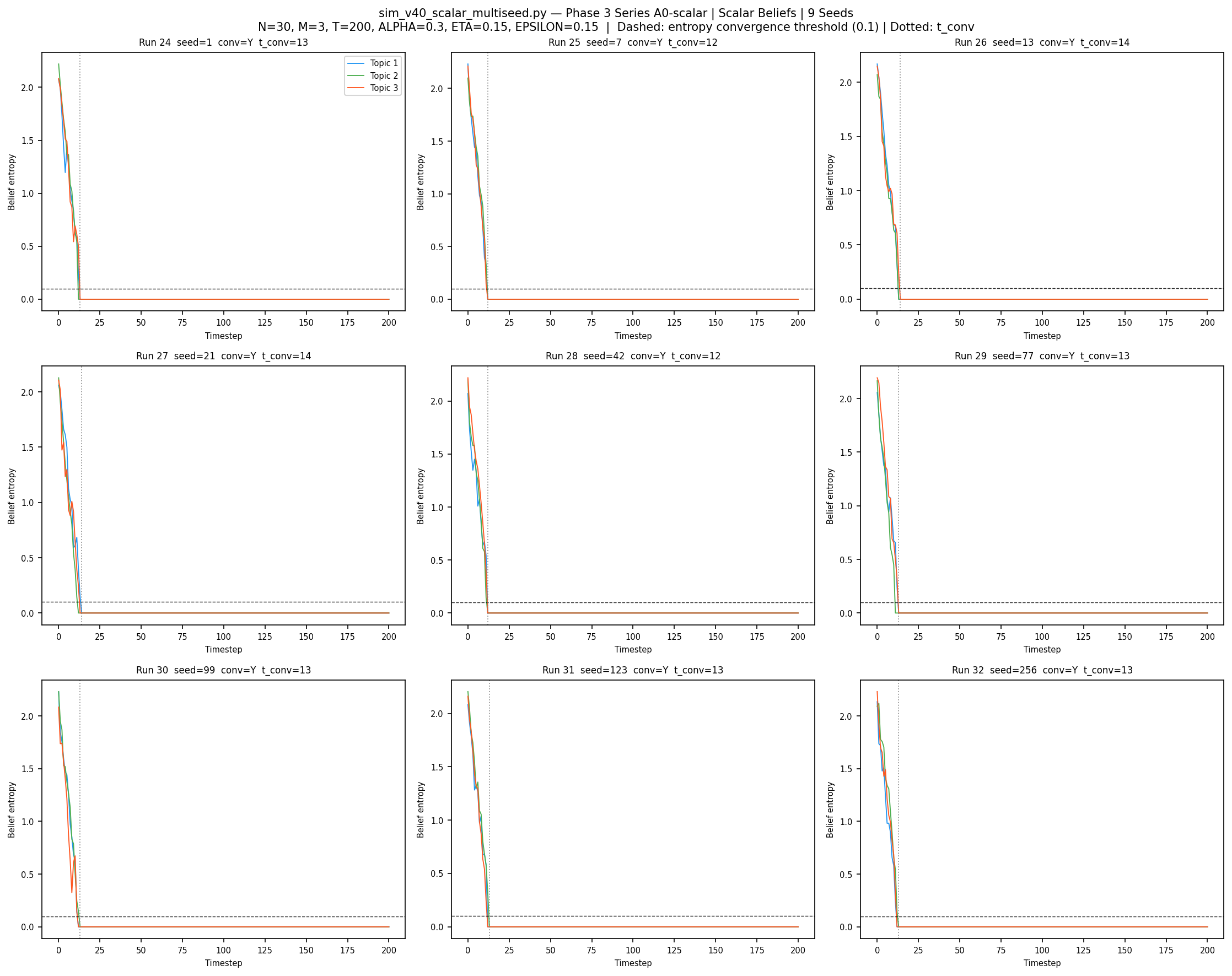}
\caption{Series A0-scalar results (Runs 24--32), one panel per seed, showing convergence, ever-SME participation, and post-convergence rotation reproducing closely across all 9 seeds --- the multi-seed confirmation underlying this section's findings.}
\label{fig3}
\end{figure}

\subsection{An Isolated-Node Edge Case in the Peer-Consistency Metric}

Run 5 showed nodes 16, 45, 85 at 97--100\% SME dominance. Three interventions (Runs 7--9) falsified degree-variance hypothesis. Cause: degree-zero nodes. $\Delta_i = 0$ maps to 1.0 under min-max normalisation --- non-participation rewarded as perfect expertise. Two-guard fix in Run 10 restored healthy rotation (top-holder 28--38\%). Required for any graph admitting isolated nodes.

\begin{figure}[htbp]
\centering
\includegraphics[width=0.95\textwidth]{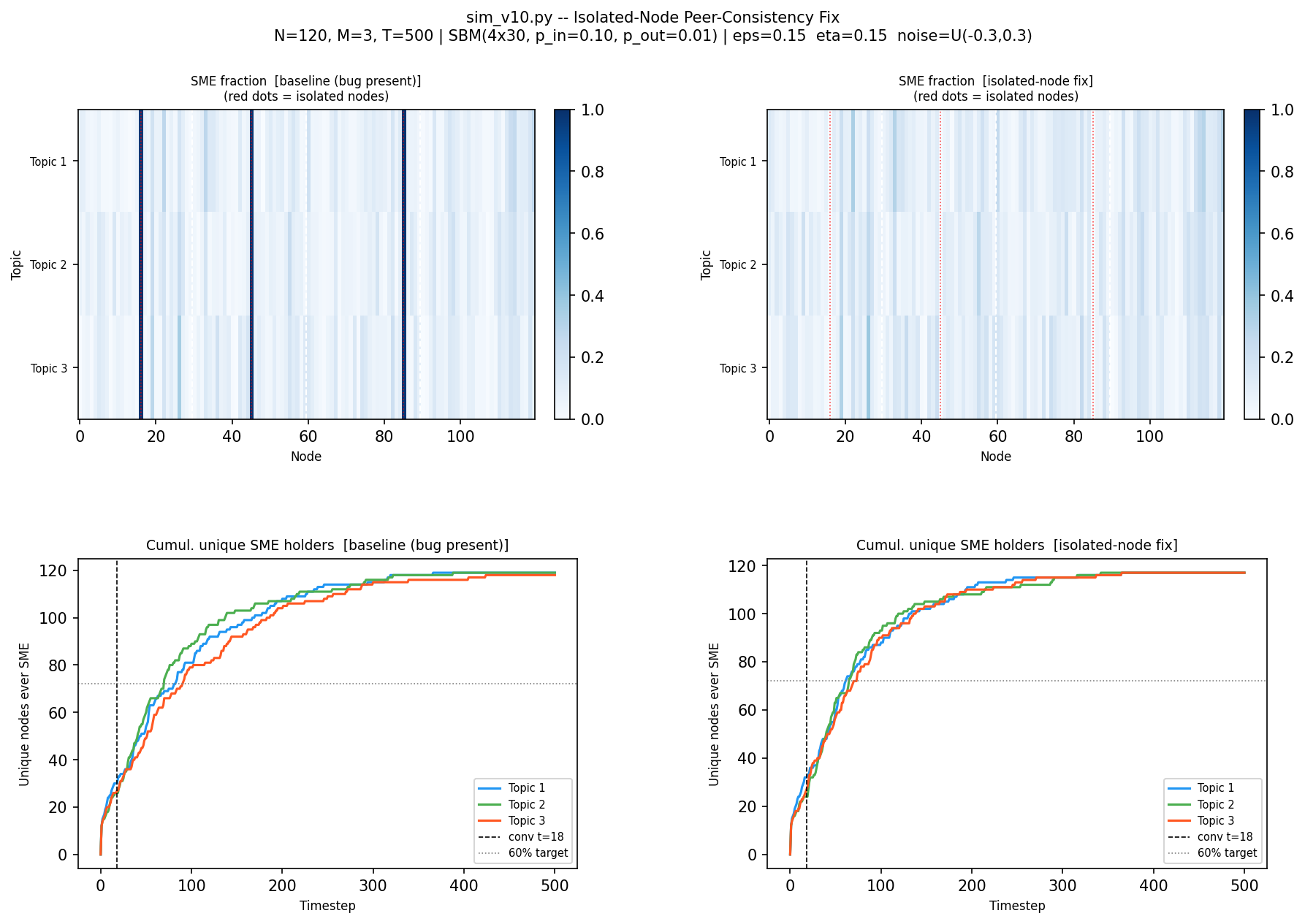}
\caption{Before/after comparison from Run 10 showing the effect of the two-guard fix for degree-zero nodes. Before the fix, isolated nodes are spuriously rewarded as perfect experts; after the fix, healthy rotation is restored (top-holder 28--38\%).}
\label{fig4}
\end{figure}

\subsection{Belief Convergence Is a Structural Property of the Update Equation}

Runs 1--3: tightening filter, increasing ETA, widening noise range --- each had zero effect on scalar consensus. Innovation noise is independent per agent; social pull is correlated across the neighbourhood. Once beliefs cluster, consensus is irreversible under the current update equation.

\subsection{Scale Invariance at N=10,000}

Run 6: 9,996--9,998/10,000 agents ever-SME. Post-convergence rotation 81.3--81.5\% --- stronger than N=120 (73--78\%). Sparse CSR implementation: peak memory below 300MB, T=500 in ~10 minutes.

\begin{figure}[htbp]
\centering
\includegraphics[width=0.95\textwidth]{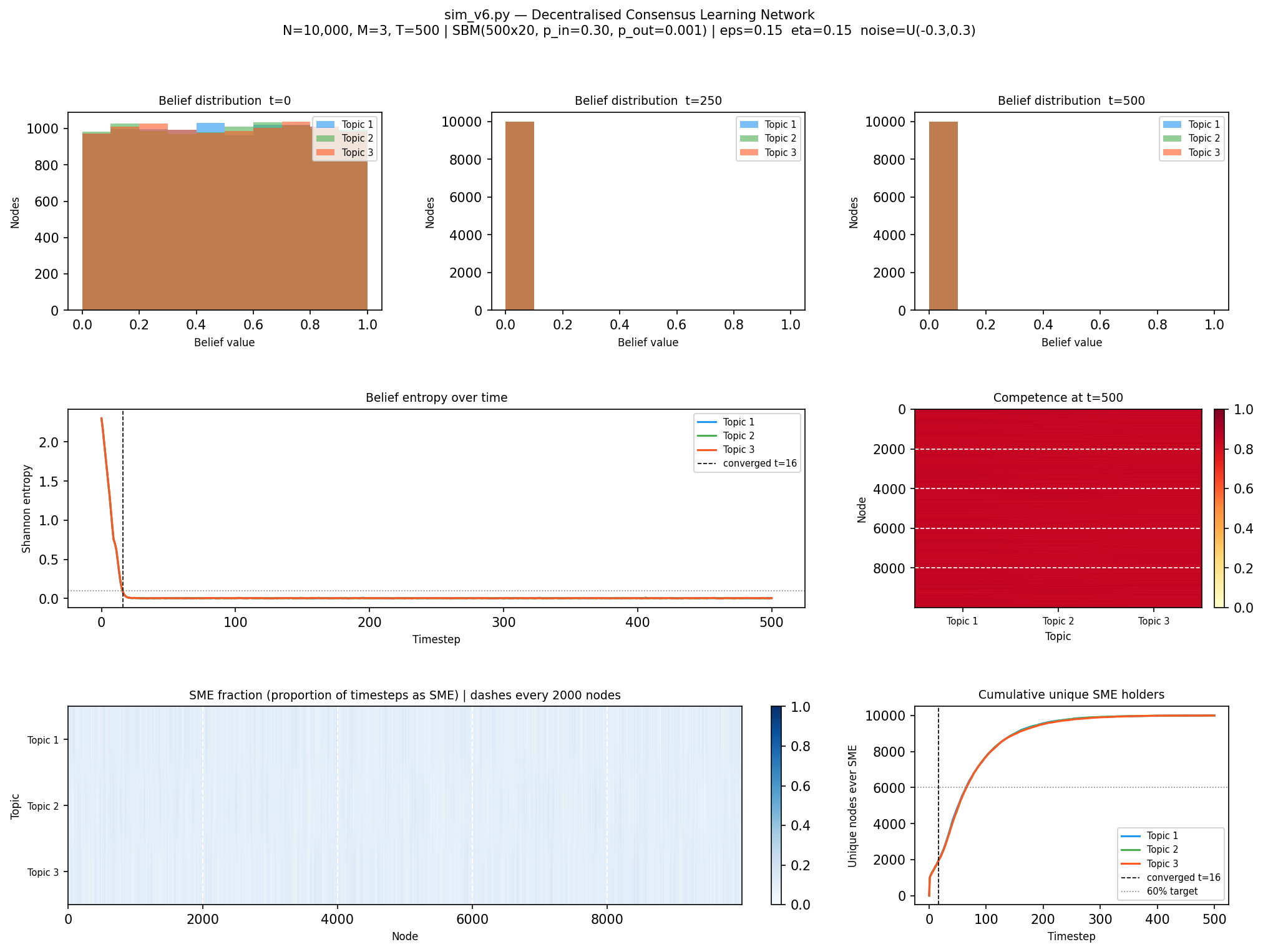}
\caption{Run 6 results at N=10,000, confirming scale invariance of SME rotation --- 9,996--9,998/10,000 agents ever-SME, with post-convergence rotation (81.3--81.5\%) strengthening relative to N=120.}
\label{fig5}
\end{figure}

\subsection{Phase 2: Vector Beliefs and the $\mathbb{R}^5$ Geometry Effect}

Runs 11--14: D=5 unit vectors. Run 11 (uniform init) converged at t=14 --- initial cosim $\approx$ 0.75 gave social pull an immediate head start. Run 12 (Gaussian init, $\kappa$=0.5) froze --- BC filter excluded all neighbour pairs. Run 13 ($\kappa$=0.0) restored learning. Run 14 (T=1000) showed slow convergence with 93--100\% ever-SME throughout.

Root cause: $\approx 0.1\sqrt{5} \approx 0.224$ angular perturbation per step, 4 orthogonal noise directions, short social centroid during early convergence. Result: dramatically extended timescales --- characterised fully in Phase 3.

\begin{figure}[htbp]
\centering
\includegraphics[width=0.95\textwidth]{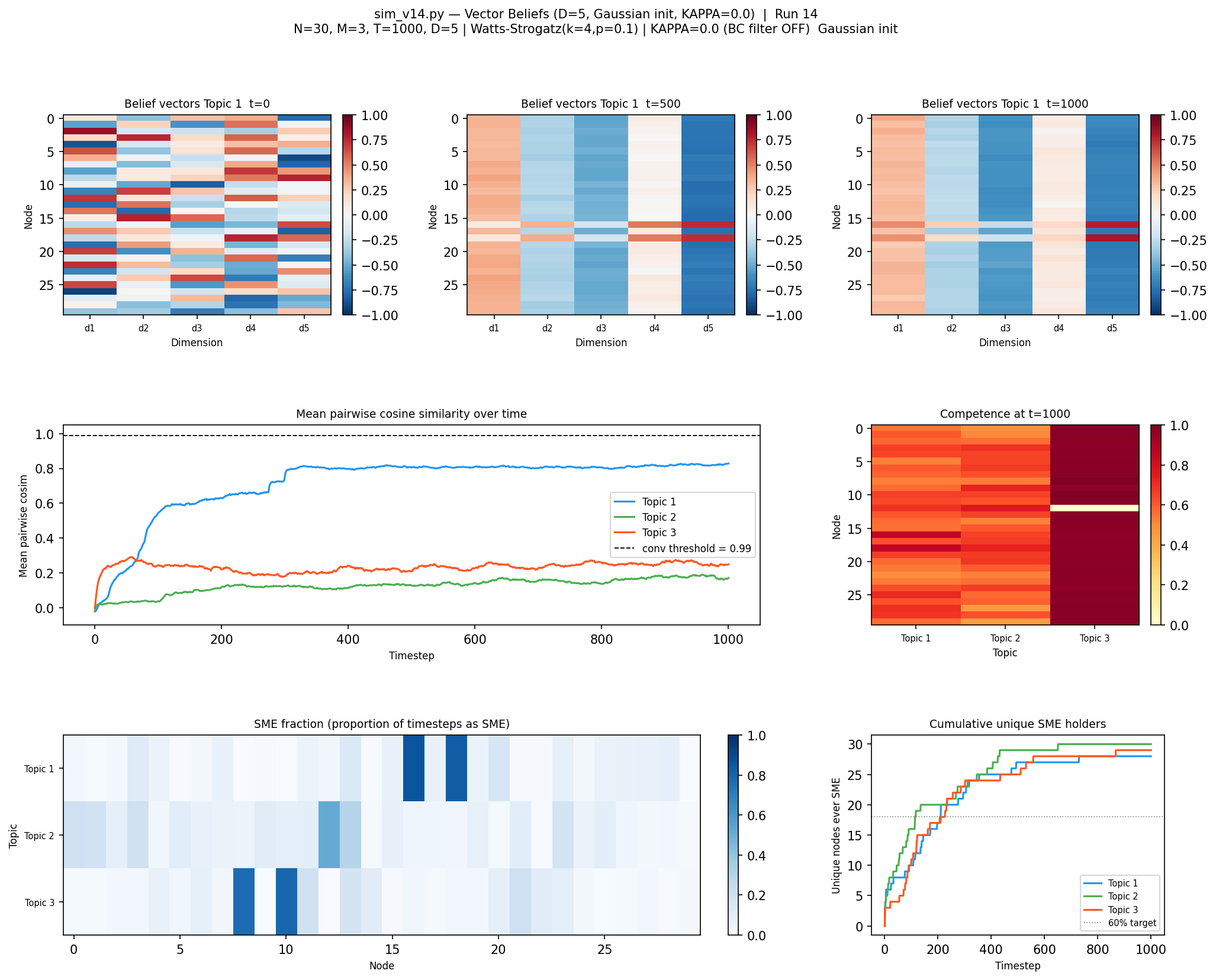}
\caption{Run 14 (T=1000) showing slow convergence in the D=5 vector-belief setting, with 93--100\% ever-SME participation maintained throughout --- the culmination of the Run 11$\to$14 progression described in this section.}
\label{fig6}
\end{figure}

\subsection{Phase 3: Convergence Timescale Heterogeneity and Cascade Dynamics}

\textbf{Series A0 (Runs 15--23, 9 seeds, T=500):} Only 1/9 seeds showed full diversity at T=500. cosim std 0.285--0.377. Within a single seed, topics diverge simultaneously --- seed 13 has Topic 1 at cosim > 0.99, Topic 3 at 0.967 (near-converged), while Topic 2 stays at 0.236. Convergence outcome per topic is determined by initial geometry, not parameters.

\begin{figure}[htbp]
\centering
\includegraphics[width=0.95\textwidth]{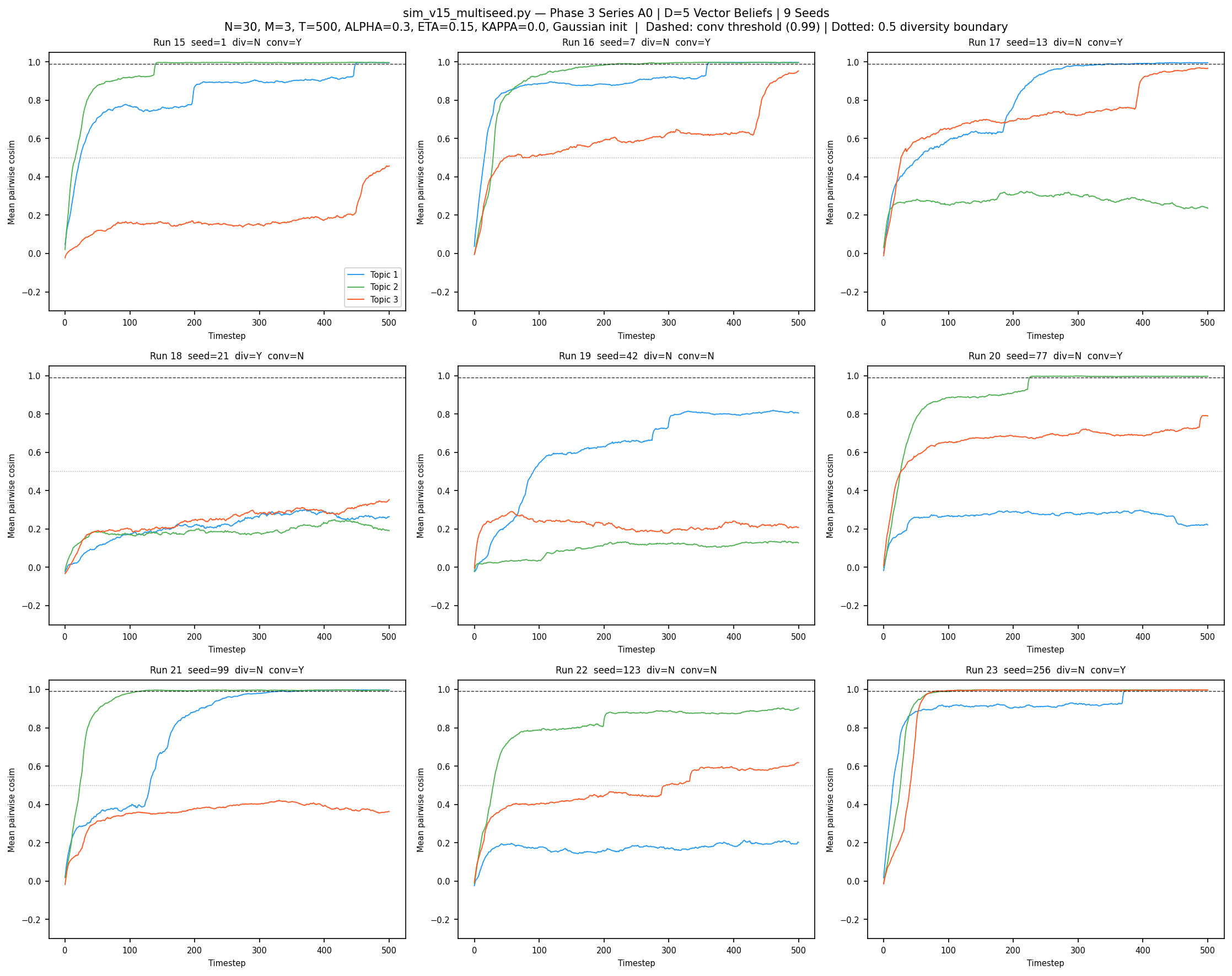}
\caption{Series A0 results (Runs 15--23), one panel per seed, showing the geometry-dependent convergence variability underlying the "1/9 seeds fully diverse at T=500" finding above.}
\label{fig7}
\end{figure}

\textbf{Series A1 (Runs 33--50, ETA sweep):} ETA 0.01--0.15 produced identical convergence classification per seed at T=500. Lower ETA paradoxically slows convergence by increasing self-retention weight. ETA is a timescale control, not a regime control.

\begin{figure}[htbp]
\centering
\includegraphics[width=0.95\textwidth]{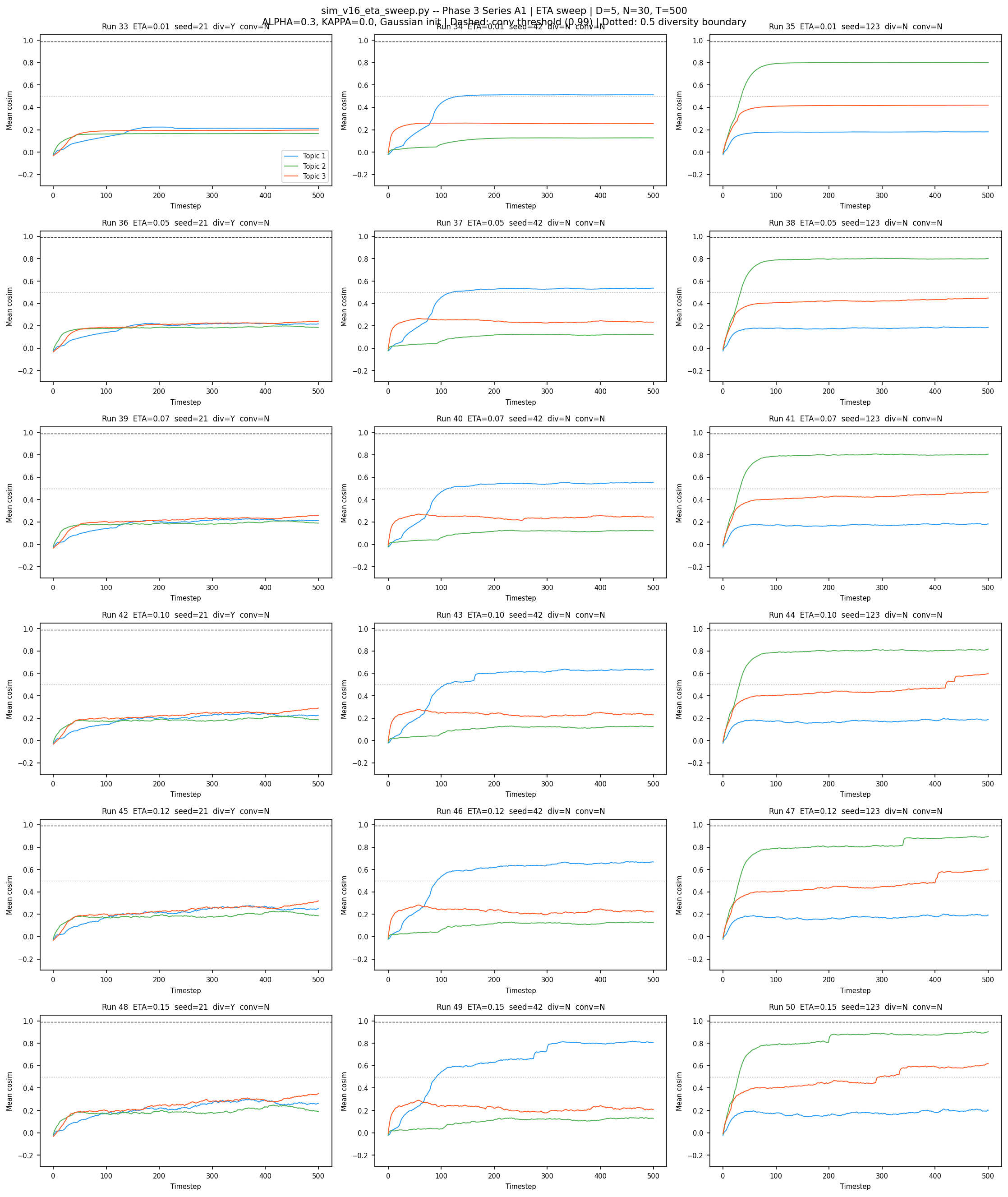}
\caption{Series A1 ETA sweep results (Runs 33--50), showing identical convergence classification across all tested ETA values per seed --- the basis for "ETA is a timescale control, not a regime control."}
\label{fig8}
\end{figure}

\textbf{Series A1 extended (Runs 51--53, T=5000):} Seed 21 (fully diverse at T=500) converged completely by T=5000. Diversity was slow convergence, not permanent.

\begin{figure}[htbp]
\centering
\includegraphics[width=0.95\textwidth]{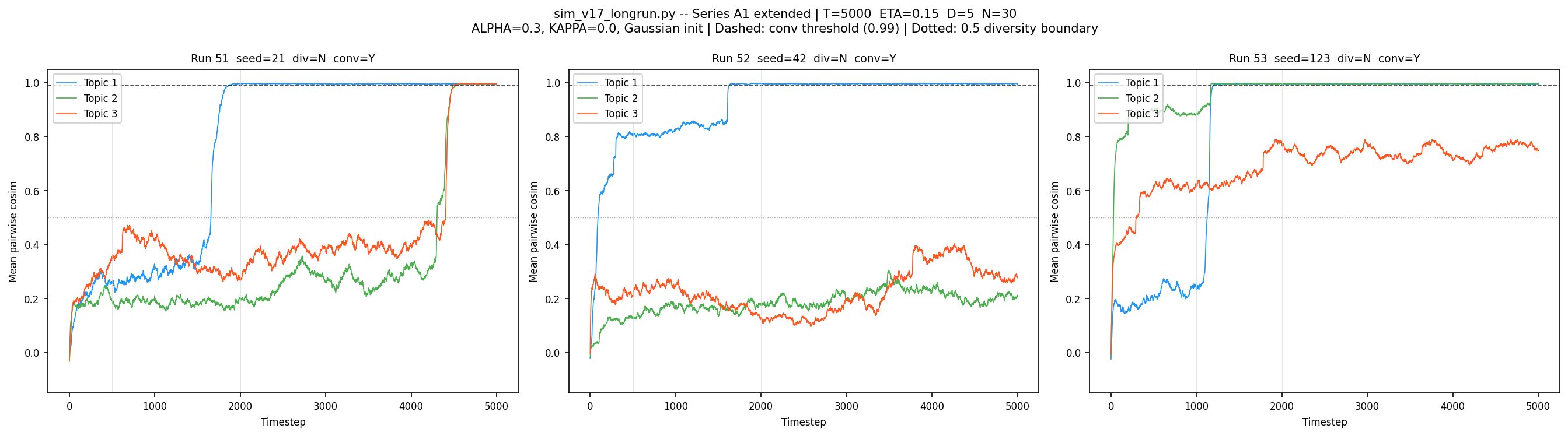}
\caption{Series A1 extended results (Runs 51--53), showing seed 21's full convergence by T=5000 --- confirming the apparent diversity at T=500 was slow convergence, not a stable state.}
\label{fig9}
\end{figure}

\textbf{Run 54 (T=20,000, seed 42):} Topic 1 at t=1,626; Topic 3 at t=11,299; Topic 2 at t=13,574. Topics 2 and 3 wandered in cosim 0.13--0.40 for ~10,000 steps before near-vertical cascade. Tipping order unpredictable from pre-cascade trajectories. Ever-SME 30/30 on all topics.

\textbf{Three-part finding:}
\begin{enumerate}
\item \textit{Timescale heterogeneity.} 3--4 orders of magnitude longer than scalar, driven by initial angular geometry.
\item \textit{Independence.} Topics converge on independent schedules on the same graph.
\item \textit{Cascade non-linearity.} Near-vertical phase transition, not gradual accumulation. Tipping time and order not predictable from pre-cascade trajectory.
\end{enumerate}

\textbf{Cluster-level idea propagation:} Topics that do not tip within operational horizon T are dropped by that cluster. The same topic may cascade in one cluster and never tip in another --- natural selective propagation from geometry alone.

\begin{figure}[htbp]
\centering
\includegraphics[width=0.95\textwidth]{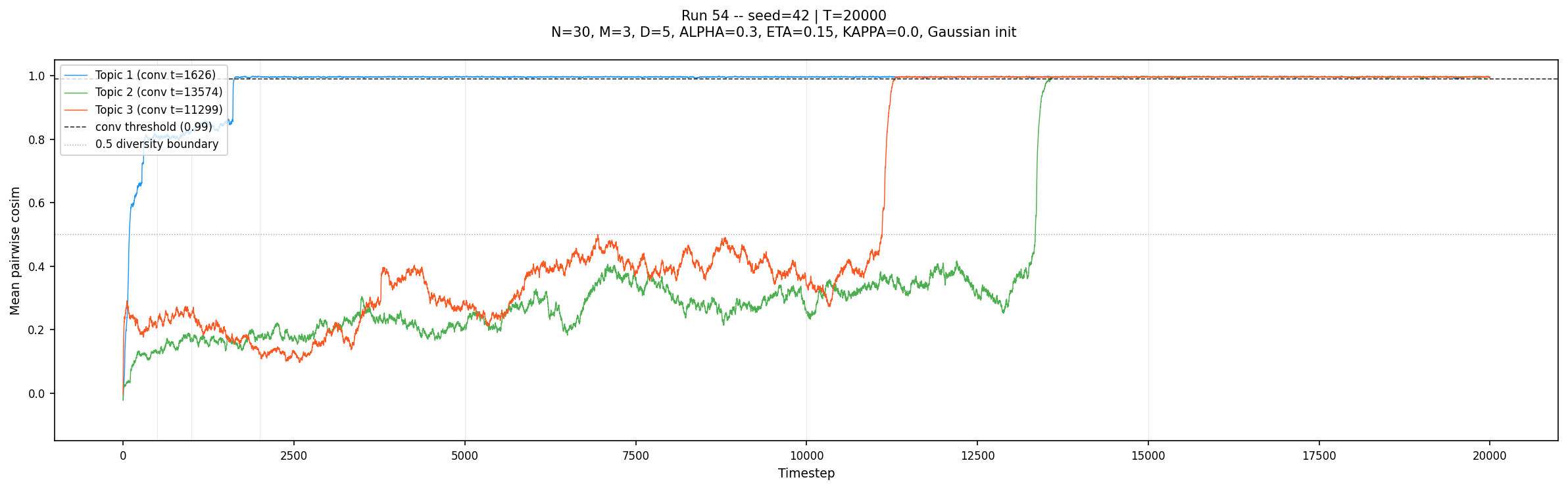}
\caption{Run 54 results, showing the near-vertical phase transitions and independent per-topic tipping schedules described above. Regenerated in Session 49 with a correctly rendered "Run 54" title --- the previously-embedded file for this slot rendered its own title as "Run 57" (from a validation re-run of this same deterministic configuration, also used as Series A2's D=5 checkpoint in Section 4.8) despite being cited here as Run 54; both are confirmed byte-identical in underlying results, so this is a labeling correction, not a data correction.}
\label{fig10}
\end{figure}

\subsection{Phase 3 Series A2: Dimensionality Sweep and Phase Structure}

Runs 55--59, D $\in \{1, 2, 5, 10, 20\}$, seed 42, ETA=0.15.

\textbf{D=1 --- Frozen (glassy):} Unit vector in 1D is {+1, -1}. Self-retention weight 0.55 prevents any sign flip. Complete immobility for all 500 steps. Oligarchy from freeze.

\textbf{D=2 --- Partial convergence:} Topic 1 cascaded at t=2,388; Topics 2 and 3 unconverged at T=20,000. Partially stuck --- one orthogonal direction limits escape routes.

\textbf{D=5 --- Cascade convergence:} Long pre-cascade wandering, independent per-topic tipping, near-vertical cascades, timescales 1,626--13,574. 30/30 ever-SME. Operational sweet spot.

\textbf{D=10 --- Non-monotone fast:} Topics 2 and 3 converged at t=109 and t=177; Topic 1 at t=6,563. Per-topic ordering reversed vs D=5 --- geometry confound dominant. Top-holder 0.28--0.31 --- healthy.

\textbf{D=20 --- Noise-floor plateau:} cosim $\approx$ 0.970--0.985 within t$\approx$5,000, stable for remaining 95,000 steps. Dynamic equilibrium: $\sigma\sqrt{20} \approx 0.447$ angular perturbation balances social pull. Top-holder 0.40-0.41.

\textbf{Geometry confound:} \texttt{rng.standard\_normal((N, M, D))} draws different values for each D --- initial angular configurations are non-comparable across D. Regime identities are robust (categorical differences); within-regime timescale comparisons across D are not attributable to dimensionality alone.

\textbf{D sweep summary:}

\begin{center}
\small
\begin{tabular}{lllll}
\toprule
D & Regime & t\_conv range & Top holder & Notes \\
\midrule
1 & Frozen & --- & 0.68--1.00 & Discrete phase space \\
2 & Partial & 2,388 to >20,000 & 0.18--0.34 & 2/3 stuck \\
5 & Cascade & 1,626--13,574 & ~0.30 & Operational sweet spot \\
10 & Non-monotone fast & 109--6,563 & 0.28--0.31 & Geometry confound dominant \\
20 & Noise-floor plateau & --- & 0.40--0.41 & Dynamic equilibrium cosim $\approx$ 0.970 \\
\bottomrule
\end{tabular}
\end{center}

\begin{figure}[htbp]
\centering
\includegraphics[width=0.95\textwidth]{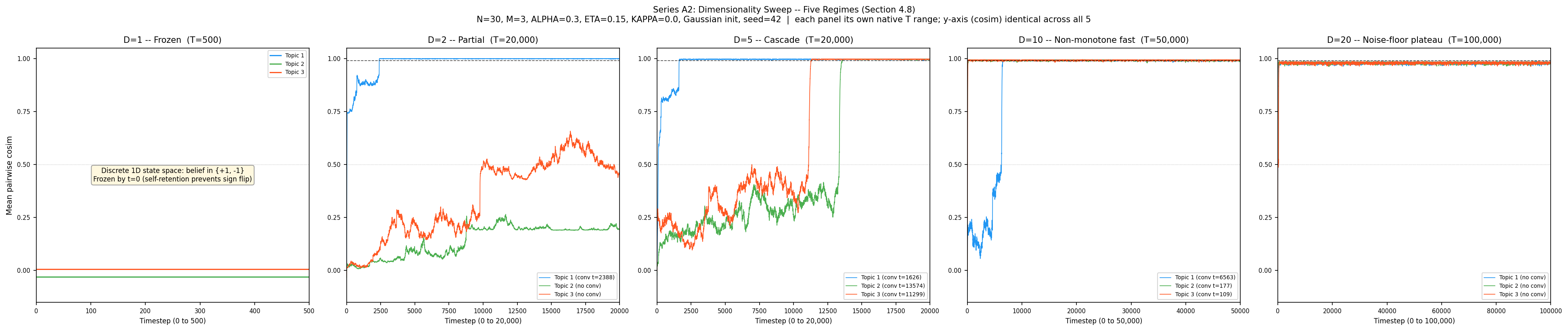}
\caption{Combined five-panel comparison across Series A2 (D=1, 2, 5, 10, 20), each panel on its own native timescale (T=500 through T=100,000) with a shared cosim y-axis, D=1's discrete {+1,-1} state space visually distinguished from the four continuous-curve panels, and each panel titled with its regime name --- replacing the earlier two-panel (D=1/D=2 only) figure with a full view of the five-regime phase structure this section describes.}
\label{fig11}
\end{figure}

\subsection{Phase 3 Series A3: High-Dimensional Equilibrium and the Diversity-Expertise Trade-off}

Series A3 extended the dimensionality investigation into the high-D regime to characterise the equilibrium behaviour and the trade-off between belief diversity and expertise concentration across that range.

\textbf{Series A3a --- High-D equilibrium sweep (Runs 60--64, D $\in \{20, 50, 100, 200, 500\}$, ETA=0.15, T=5000):}

All five D values produced the same qualitative behaviour: rapid rise to equilibrium cosim within a few hundred steps, followed by a perfectly stable plateau for the remainder of T=5000. Equilibrium is reached faster at higher D --- at D=500 the plateau is established by t$\approx$94, at D=20 by t$\approx$330. The equilibrium cosim falls monotonically with D:

\begin{center}
\small
\begin{tabular}{lllll}
\toprule
D & eq\_cosim & t\_to\_eq & minSME & maxHolder \\
\midrule
20 & 0.979 & 330 & 28/30 & 0.448 \\
50 & 0.939 & 144 & 24/30 & 0.605 \\
100 & 0.881 & 97 & 21/30 & 0.771 \\
200 & 0.769 & 90 & 17/30 & 0.919 \\
500 & 0.484 & 94 & 14/30 & 0.991 \\
\bottomrule
\end{tabular}
\end{center}

Two findings of note. First, the equilibrium cosim follows the noise scaling prediction: $\sigma\sqrt{D}$ grows monotonically, pushing the noise floor lower. Second, and critically, SME rotation degrades monotonically with D: maxHolder rises from 0.448 at D=20 to 0.991 at D=500. At D=500, one agent holds top SME status 99.1\% of timesteps despite eq\_cosim=0.484 --- the network maintains rich belief diversity but frozen expertise. High-dimensional noise does not prevent expert hierarchy; it amplifies it by making the peer-consistency signal noisy for all agents, which causes small initial competence differences to compound rather than erode.

\begin{figure}[htbp]
\centering
\includegraphics[width=0.95\textwidth]{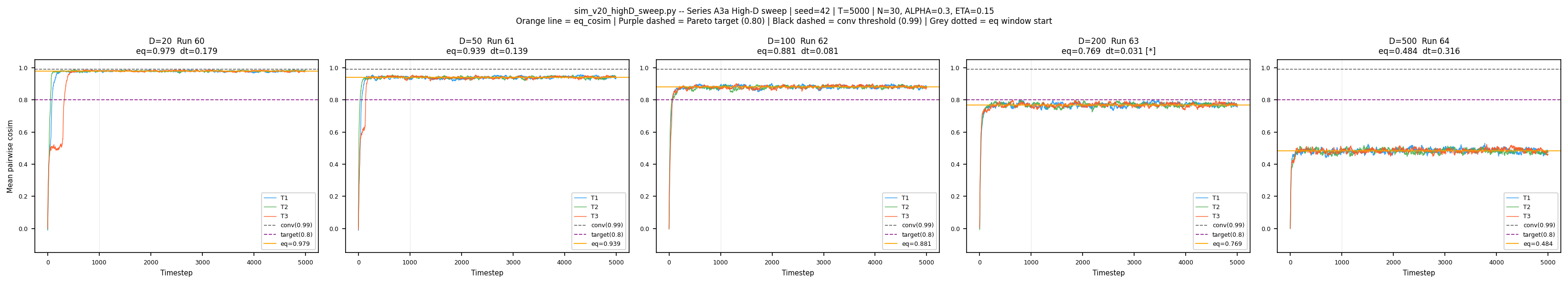}
\caption{Series A3a results (Runs 60--64) showing the monotonic fall in equilibrium cosim and the corresponding rise in maxHolder across D={20,50,100,200,500} --- the core diversity-expertise trade-off of Series A3.}
\label{fig12}
\end{figure}

\textbf{Series A3b sub-run 1 --- Fine D sweep (Runs 65--67, D $\in \{125, 150, 175\}$, ETA=0.15, T=5000):}

\begin{center}
\small
\begin{tabular}{llll}
\toprule
D & eq\_cosim & minSME & maxHolder \\
\midrule
125 & 0.850 & 19/30 & 0.833 \\
150 & 0.822 & 21/30 & 0.889 \\
175 & 0.795 & 16/30 & 0.906 \\
\bottomrule
\end{tabular}
\end{center}

At seed=42, D=175 produced eq\_cosim = 0.795, reached within approximately 79 steps and maintained indefinitely. All three topic lines converge to the same equilibrium simultaneously, with no per-topic divergence --- the high-D regime produces topic-uniform equilibria, in contrast to the cascade regime (D=5) where topics evolve on independent schedules.

\begin{figure}[htbp]
\centering
\includegraphics[width=0.95\textwidth]{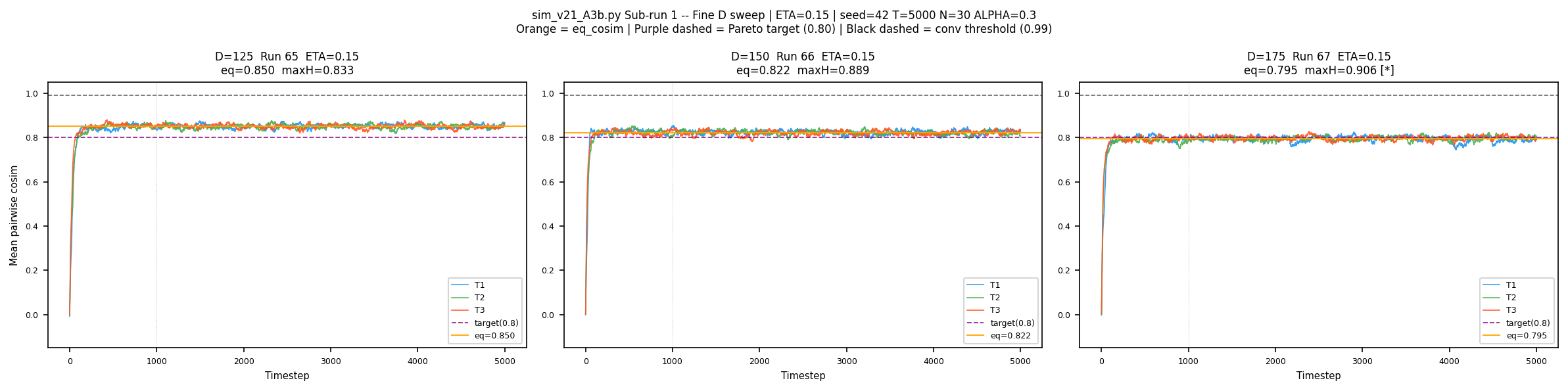}
\caption{Series A3b sub-run 1 (Runs 65--67), showing the fine D sweep across D={125,150,175} at seed=42 (eq\_cosim=0.795 at D=175). The consensus-vs-D trend shown here was later confirmed across five additional seeds (Series A3c, discussed below) as a real, consistent direction rather than a single reproducible number.}
\label{fig13}
\end{figure}

A later seed-robustness check (Series A3c, Runs 70--84) tested D=150, 175, and 200 across five additional seeds. Five of six seeds cluster tightly, landing in the ~0.76--0.83 eq\_cosim range across the tested band, with higher D correlating with lower consensus overall. This direction holds for every clustering seed across the full D=150--200 span, with one mild exception at an intermediate step: seed 7 rises slightly from D=150 to D=175 (0.8159 $\to$ 0.8209) before falling again at D=200 (0.7811) --- consistent with ordinary noise around a real downward trend rather than a break in it. The sixth seed (11) is a wider but structurally-explained outlier: its network has one node at unusually low degree (2, versus every other seed's minimum of 3) --- independently verified by directly regenerating each seed's Watts-Strogatz graph (\texttt{nx.watts\_strogatz\_graph(30, k=4, p=0.1, seed=X)}; see \texttt{degree-verification.py} in the same output folder) rather than left as an unsourced assertion --- plausibly slowing that node's pull into consensus; it is reported here as natural variation, not excluded. This variability is expected rather than a weakness --- the model's random innovation term represents new arguments entering each agent's thinking at different points in the run, and the same path-dependence shows up in real deliberation, where the same underlying topic can converge differently depending on how debate actually unfolds. No single D value within this band is presented as a fixed target; the trend itself --- not a specific equilibrium number --- is the finding.

\begin{figure}[htbp]
\centering
\includegraphics[width=0.95\textwidth]{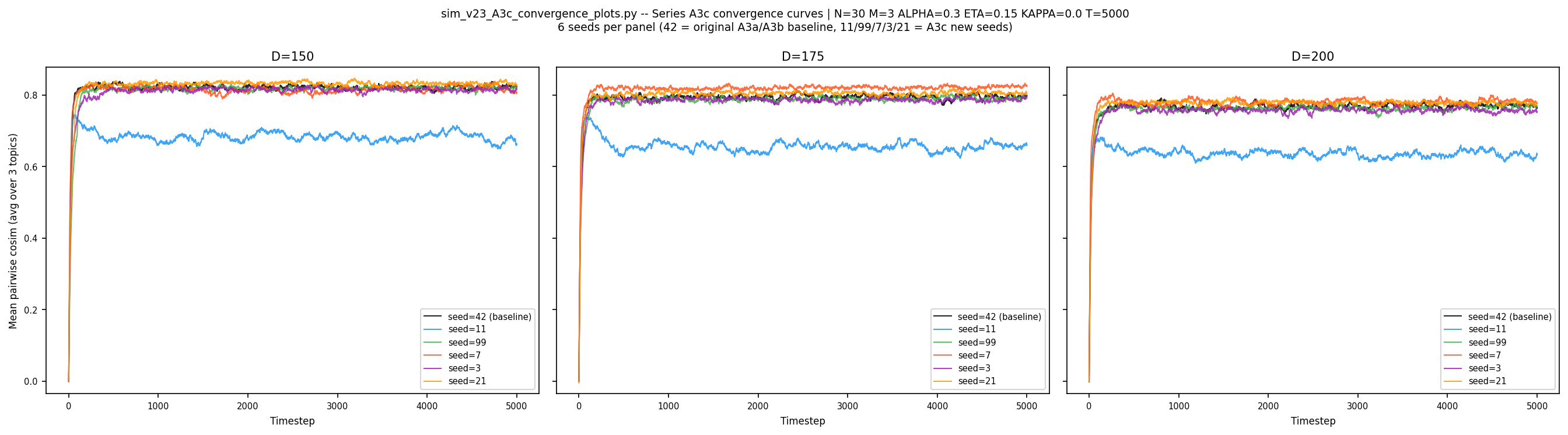}
\caption{Series A3c results (Runs 70--84), one panel per seed, showing five seeds clustering tightly across the D=150--200 band while seed 11 lands as a wider, structurally-explained outlier --- the seed-robustness evidence underlying the eq\_cosim $\approx$ 0.76--0.83 range reported above.}
\label{fig14}
\end{figure}

\textbf{Series A3b sub-run 2 --- ETA sensitivity at D=150 (Runs 68--69, ETA $\in \{0.05, 0.10\}$, D=150, T=5000):}

\begin{center}
\small
\begin{tabular}{lll}
\toprule
ETA & eq\_cosim & maxHolder \\
\midrule
0.05 & 0.982 & 0.867 \\
0.10 & 0.924 & 0.884 \\
0.15 & 0.822 & 0.889 \\
\bottomrule
\end{tabular}
\end{center}

The ETA hypothesis is falsified. Reducing ETA from 0.15 to 0.05 shifts eq\_cosim from 0.822 to 0.982 --- collapsing belief diversity almost entirely --- while maxHolder barely moves (0.889 $\to$ 0.867). ETA and D are not independent design levers. D is the primary control for equilibrium belief diversity. ETA is a secondary timescale modulator that cannot recover rotation health at high D. The expertise concentration is dimensionality-driven, not noise-driven.

\begin{figure}[htbp]
\centering
\includegraphics[width=0.95\textwidth]{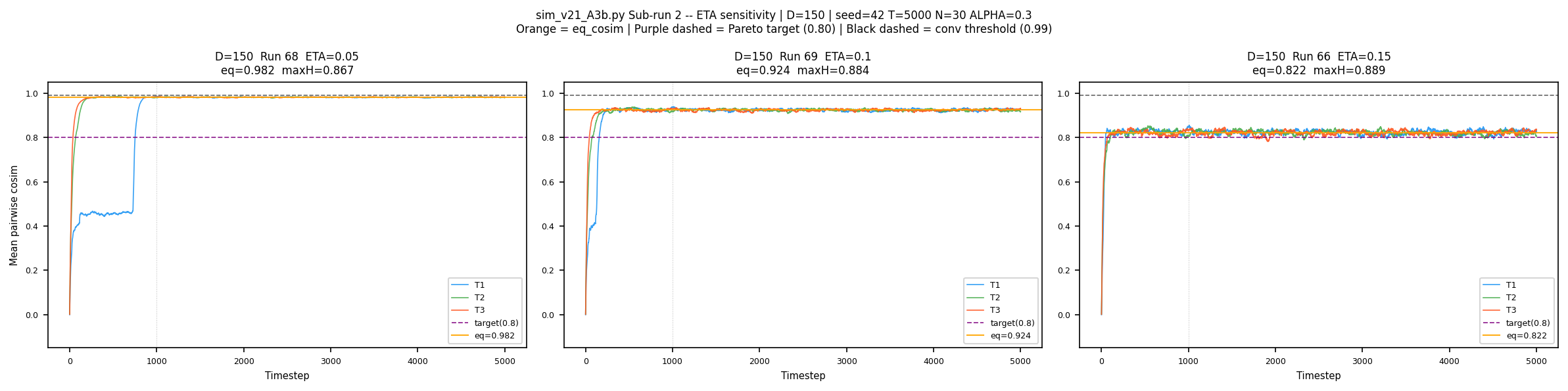}
\caption{Series A3b sub-run 2 results (Runs 68--69), showing eq\_cosim's strong response to ETA against maxHolder's near-flat response --- the basis for the ETA falsification described above.}
\label{fig15}
\end{figure}

\textbf{The diversity-expertise trade-off:}

Series A3 reveals a monotone, unavoidable trade-off as D increases: belief diversity increases (eq\_cosim decreases toward the Pareto zone) while SME expertise concentration also increases (maxHolder rises toward oligarchy). No (D, ETA) combination tested achieves both eq\_cosim $\approx$ 0.80 and maxHolder < 0.50 simultaneously.

\textbf{Reinterpretation of high-D expertise concentration:}

This trade-off is not necessarily a failure of the rotation mechanism --- it is the expected signature of the mechanism correctly identifying and locking onto genuine competence at equilibrium. Early in a run, competence has not yet sorted itself out: SME status shifts between several competing agents as the network tests candidates before committing, which looks like a genuine power struggle. This volatility is not a problem to explain away --- it is the mechanism doing its job. Over time, as the peer-consistency process converges, one agent's position solidifies and the network settles: at high D, the agent holding SME status most of the time is the agent whose high-dimensional belief vector is most consistently aligned with its neighbours' --- tracking a complex multi-dimensional consensus direction across all dimensions simultaneously. In a near-consensus state with meaningful residual diversity, maintaining that alignment is genuinely non-trivial. This is not accidental advantage from initial geometry --- as it is during the pre-cascade transient at D=5 --- but earned, sustained alignment tracking in a high-complexity belief space. In real knowledge communities, the agent who most consistently tracks a complex multi-dimensional consensus direction is precisely what we call an expert. The model's competence mechanism is correctly identifying that agent.

The distinction from pathological oligarchy (the isolated-node boundary condition in Section 4.3) is important: there, non-participation was rewarded as perfect expertise. Here, sustained participation and consistent alignment tracking earns expertise. The mechanism is operating as designed; the high-D regime simply reveals a regime where one agent is genuinely better at the task the mechanism is measuring.

Exactly which agent ends up dominant, and how completely, is not itself something a specific D value within the high-D range controls. Once the network is already in high-D territory, the peer-consistency correction signal is spread thin across many dimensions and becomes noisier --- the Series A3c seed-robustness check (above) found that maxHolder's exact value at a fixed D varies considerably across seeds, dominated by initial conditions (who started with a slightly higher competence edge, network wiring) rather than by fine adjustments to D. That seed-to-seed variability is not evidence against the mechanism; it is a natural consequence of weaker self-correction at high dimensionality, operating on whatever small initial advantage happened to exist.

\textbf{Practical design implication:} For applications where both belief diversity and rotation health are required simultaneously, D=50 is the recommended operating point: eq\_cosim=0.939 (strong broad agreement), maxHolder=0.605 (tolerable rotation). For applications where diversity is the primary objective and stable expertise is acceptable or desirable, the D=150--200 band delivers consistently high diversity with settled expertise concentration; no single D value within that band is a more reliable choice than another, since the exact landing point depends on seed-level path-dependence rather than the specific D chosen.

\section{Discussion}

\subsection{Implications for the Core Hypothesis}

The central claim --- that meaningful expertise rotation emerges from peer-consensus dynamics without centralised reward --- is supported robustly across 84 runs, three scales, two topologies, and eight belief dimensionalities. SME rotation was observed in every primary run, and Series A0-scalar's 9-seed check (Section 4.2a) confirms this is not a seed-42 artefact for the scalar case specifically. Post-convergence rotation strengthens with scale. Run 54 confirmed 30/30 ever-SME at T=20,000. Series A2 confirmed 30/30 ever-SME at D=20 over T=100,000. Series A3 confirmed strong ever-SME participation across D={20,50,100,200,500} at T=5000 --- ranging from 28/30 at D=20 down to 14/30 at D=500 as maxHolder concentration increases with D --- the competence mechanism remains active across all regimes tested, even as fewer agents ever reach SME status at the highest D values.

The diversity-expertise trade-off identified in Series A3 adds nuance to this claim. Rotation health degrades with D --- the model does not simultaneously optimise for belief diversity and distributed expertise at high dimensionality. However, the reinterpretation of Section 4.9 is important: at high D, the concentration of expertise is not a malfunction but a signal. The agent who most consistently tracks a complex, high-dimensional consensus direction in a state of meaningful residual diversity is correctly identified as the expert. The model is doing what it was designed to do; the high-D regime reveals that "expertise" and "distributed expertise" are not the same thing, and that the appropriate balance between them is a design choice rather than a universal optimum.

The high-D operating zone identified across Series A3 (D$\approx$150--200, confirmed as a consistent trend across six seeds in Series A3c) is among the most practically actionable findings of the series. A network operating in this zone requires no external diversity-preservation mechanism --- the geometry produces strong consensus alongside meaningful, persistent diversity as a stable outcome, reached on the order of 100 steps (observed range 25--137 across the tested seeds) and maintained indefinitely, without needing to target one exact D value. This stands in direct contrast to centralised reward architectures, which require explicit regularisation, dropout, or diversity bonuses to avoid collapse. Here diversity is structural, not engineered.

The Series A3 ETA falsification also sharpens the architectural parallel. A centralised reward system operating at high embedding dimensionality (D=768+) with low innovation noise (ETA$\approx$0) would sit in the upper-left corner of the (D, ETA) space: high D, low ETA, eq\_cosim approaching 1.0, full expertise concentration. This is precisely the structure of a finetuned language model --- extremely high-dimensional belief space, near-zero innovation relative to the training attractor, near-consensus on all topics. The Series A3 results show that this is not a tuning failure but a structural consequence of operating at high D with low ETA. Moving toward the Pareto zone requires increasing D while maintaining ETA --- or reducing effective D through projection --- neither of which is straightforward in a centralised training paradigm.

\subsection{Limitations}

\textbf{Scale --- partially addressed.} N=10,000 confirmed for scalar beliefs. High-D vector belief runs remain at N=30.

\textbf{D sweep --- completed through Series A3, geometry confound noted.} The controlled D sweep (fixed per-topic initialisation projected onto each D) remains as future work to isolate pure dimensionality effects.

\textbf{Diversity-expertise trade-off --- characterised, not resolved.} Series A3 established the trade-off as monotone and ETA-independent. Whether alternative competence update mechanisms (e.g. anti-persistence terms, neighbourhood diversity bonuses) can decouple D from maxHolder remains open.

\textbf{No idea objects.} Cluster-level propagation inferred from per-topic convergence dynamics. Explicit epidemic-style propagation would allow direct empirical testing.

\textbf{Single seed --- addressed for Phase 1 (scalar) and the D=5 vector case; still open elsewhere.} Series A0 tested 9 seeds at D=5 (Section 4.7); Series A0-scalar (Section 4.2a) closes the equivalent gap for Phase 1's scalar case. Series A3 uses seed 42 throughout across its full D range; multi-seed characterisation there remains deferred pending controlled initialisation work.

\textbf{Dynamic $\alpha$ not modelled.} Topology-aware dynamic $\alpha$ --- stronger within clusters, weaker across bridge edges --- remains a natural Phase 4 direction.

\textbf{Anti-oligarchy at degree-skew scale untested.} $\mu$ trust penalty untested against preferential attachment graphs.

\textbf{Entrenched-expertise displacement by a genuine competing idea --- untested.} This is distinct from the single-seed limitation above, which concerns whether a \textit{result} generalises across seeds. This concerns whether the \textit{mechanism} itself can be legitimately challenged: the model has not been tested for whether an entrenched SME's status can be displaced by a real, structured competing idea, as opposed to the model's existing random "innovation" term, which has no persistence and no coherent content and so cannot represent a genuine alternative argument the way a real new idea would. This is an open question the current architecture cannot answer, not a demonstrated flaw: does the network correctly update its expert assignment when a better idea appears, or does entrenchment resist it regardless of merit? This motivates a planned Phase 4 prototype --- a deliberate "shock belief" experiment that injects a structured competing idea rather than random noise --- though no Phase 4 results exist yet.

\subsection{Remaining Directions}

Three simulation directions remain open.

First, a controlled D sweep with fixed per-topic initial configurations projected onto each D, isolating the pure dimensionality effect from the geometry confound.

Second, an alternative competence update study at high D --- testing whether anti-persistence terms or neighbourhood diversity bonuses can recover rotation health without collapsing eq\_cosim, potentially breaking the diversity-expertise trade-off.

Third, implementing discrete idea propagation --- idea objects diffusing through the trust graph with quality-weighted probability --- to empirically test the cluster-level propagation hypothesis.

The primary direction beyond simulation is Workstream B: a multi-agent prototype where Claude API instances operate as nodes, sentence embeddings projected to D=5 serve as belief vectors, and the competence update equation governs trust weighting across rounds. The D sweep and A3 results provide two design parameters: project to D=5 for cascade dynamics with healthy rotation, or project to D=50 for the plateau regime with tolerable rotation and stronger diversity. Both are now principled choices rather than guesses.

\section{Conclusion}

We presented a decentralised consensus-based learning network and demonstrated through eighty-four agent-based simulation runs that SME rotation is robust (90--100\% ever-SME across all primary runs and a 9-seed robustness check for the scalar case), persistent (67--86\%, mean $\approx$77\%, post-convergence rotation confirmed across 9 seeds for the scalar case, 30/30 participation confirmed at T=20,000 and T=100,000 in vector runs), topology-invariant, and scale-invariant.

Phase 3 Series A0 and A1 established convergence timescale heterogeneity with cascade dynamics as the definitive characterisation of vector belief behaviour at D=5. Series A2 extended this to a five-regime phase structure across D={1,2,5,10,20} with genuine categorical differences between regimes. Series A0-scalar (Section 4.2a) closed the equivalent single-seed gap for the original scalar-belief result: 9 independent seeds reproduce both the qualitative rotation pattern and, closely, the quantitative range first observed at seed 42 alone.

Series A3 characterised the high-dimensional equilibrium regime across D={20,50,100,125,150,175,200,500} and produced three findings. First, equilibrium cosim falls monotonically with D following the $\sigma\sqrt{D}$ noise scaling; within the D=150--200 band, a later six-seed check (Series A3c) confirmed this as a real, consistent trend --- five of six seeds cluster in the eq\_cosim $\approx$ 0.76--0.83 range, with one seed (7) showing a mild non-monotonic step at an intermediate D value and the sixth (seed 11) a wider but structurally-explained and independently-verified outlier --- rather than a single reproducible equilibrium value at one D. Second, SME expertise concentration rises monotonically with D --- an unavoidable trade-off between belief diversity and rotation health that ETA cannot modulate, falsifying the hypothesis that D and ETA are independent design levers. Third, the expertise concentration at high D is reinterpreted as correct behaviour: the agent consistently tracking a complex, high-dimensional consensus direction in a state of meaningful residual diversity is correctly identified as the expert by the peer-consistency mechanism. This is earned alignment tracking, not accidental advantage --- and the seed-to-seed variability in exactly how concentrated that expertise becomes is itself expected, since the peer-consistency correction signal grows noisier at high D and initial conditions dominate the precise outcome.

The high-D operating zone (D$\approx$150--200) --- strong consensus alongside meaningful, persistent diversity, reached on the order of 100 steps (observed range 25--137 across tested seeds) and maintained indefinitely without external intervention --- is among the most practically actionable findings of the series. It demonstrates that belief diversity and collective alignment are not mutually exclusive in the vector belief network, and that their coexistence is a structural property of the geometry rather than a designed feature. For system builders, D=50 (eq\_cosim=0.939, maxHolder=0.605) is the recommended operating point when both diversity and rotation health are required simultaneously. The D=150--200 band is appropriate when persistent diversity with stable expertise is the design target; no single value within that band should be treated as a tuned optimum, since the exact equilibrium a given run reaches depends on seed-level path-dependence rather than the specific D chosen.

Together these findings formalise the selective, geometry-driven idea propagation of real knowledge ecosystems and provide a principled architectural complement --- and a diagnostic framework --- for the innovation ceiling of centralised reward systems.

\section{Code Availability}

All simulation code, run scripts, and output data underlying this paper are available at \href{https://github.com/SomeLameCode/decentralised-consensus-networks}{https://github.com/SomeLameCode/decentralised-consensus-networks}, released under the MIT license. The repository is organised into \texttt{phase-1/}, \texttt{phase-2/}, and \texttt{phase-3/} folders corresponding directly to the three phases described in Section 3, each with its own README documenting every run's exact parameters and purpose. Every run seeds both its graph generation and its random number generator, so any result reported in this paper can be reproduced exactly by re-running the corresponding script.

\section{References}

Deffuant, G., Neau, D., Amblard, F., \& Weisbuch, G. (2000). Mixing beliefs among interacting agents. \textit{Advances in Complex Systems}, 3(01n04), 87--98. https://doi.org/10.1142/S0219525900000078

DeGroot, M.H. (1974). Reaching a consensus. \textit{Journal of the American Statistical Association}, 69(345), 118--121. https://doi.org/10.1080/01621459.1974.10480137

Holland, J.H. (1995). \textit{Hidden Order: How Adaptation Builds Complexity}. Addison-Wesley (Helix Books), Reading, MA.

Watts, D.J., \& Strogatz, S.H. (1998). Collective dynamics of 'small-world' networks. \textit{Nature}, 393, 440--442.
\end{document}